\documentclass[]{ceurart}

\usepackage{listings}
\usepackage{todonotes}

\begin{document}

\copyrightyear{2023}
\copyrightclause{Copyright for this paper by its authors. Use permitted under Creative Commons License Attribution 4.0
  International (CC BY 4.0).}

\conference{2nd Edition of EvalRS: a Rounded Evaluation of Recommender Systems, August 6 - August 10, 2023, Long Beach, CA, USA}

\title{Bridging Offline-Online Evaluation with a Time-dependent and Popularity Bias-free Offline Metric for Recommenders}

\author{Petr Kasalický}[%
orcid=0000-0001-6438-366X,
email=kasalpe1@fit.cvut.cz
]
\cormark[1]
\address{Faculty of Information Technology, Czech Technical University in Prague,
  Thákurova 9, Prague, 160 00, Czech Republic}

\author{Rodrigo Alves}[%
orcid=0000-0001-7458-5281,
email=rodrigo.alves@fit.cvut.cz
]

\author{Pavel Kordík}[%
orcid=0000-0003-1433-0089,
email=pavel.kordik@fit.cvut.cz
]

\cortext[1]{Corresponding author.}

\begin{abstract}
  The evaluation of recommendation systems is a complex task. The offline and online evaluation metrics for recommender systems are ambiguous in their true objectives. The majority of recently published papers benchmark their methods using ill-posed offline evaluation methodology that often fails to predict true online performance. Because of this, the impact that academic research has on the industry is reduced. The aim of our research is to investigate and compare the online performance of offline evaluation metrics. We show that penalizing popular items and considering the time of transactions during the evaluation significantly improves our ability to choose the best recommendation model for a live recommender system. Our results, averaged over five large-size real-world live data procured from recommenders, aim to help the academic community to understand better offline evaluation and optimization criteria that are more relevant for real applications of recommender systems.
\end{abstract}


\maketitle

\section{Introduction}
The evaluation of recommendation systems is a complex task. One of the primary reasons is that several evaluation metrics represent distinct properties of a recommendation algorithm (RA). For instance, RMSE is a reliable marker of how a RA predicts a user-to-item taste \cite{ledent_orthogonal_2021}, whereas Recall reflects how well the same algorithm retrieves a relevant list of items \cite{silveira_how_2019}. Therefore, it is crucial to identify metrics that effectively evaluate and describe the target of a particular recommender system (RS).

A comprehensive search of the relevant literature revealed that the most prevalent method is \emph{offline} evaluation of static feedback, where data are collected from a real-world RS dataset \cite{ni_justifying_2019, wannigamage_steam_2020}. In offline evaluation, recommendation algorithms are trained on a training subset of data \cite{handbook-evaluating-recommender-systems}, and then their recommendations are evaluated on test data using metrics such as Recall@$k$ \cite{wang_knowledge-aware_2019}, Precision@$k$ \cite{sachdeva_sequential_2018}, NDCG@$k$ \cite{steck_embarrassingly_2019}, MAP@$k$ \cite{yang_bridging_2017} and HR@$k$ \cite{tay_latent_2018}. 

However, conventional offline evaluation of RS has significant disadvantages. For instance, the interaction observation follows a non-uniform distribution~\cite{marlin_collaborative_2007, marlin_collaborative_2009, steck_item_2011}: certain items have more interaction observations solely because the RS exposed them to more users. In addition to observation bias, standard offline evaluation techniques are agnostic to the fact that recommendation algorithms perform in a live environment where it is only possible to use interactions made in the past to predict future interactions: the commonly employed leave-one-out-cross-validation (LOOCV) method \cite{li_collaborative_2017,vinagre_evaluation_2014} does not track the user's behavior over time.

Several approaches have been proposed to address these drawbacks. Bias in the test data caused by the missing-not-at-random (MNAR) problem can be partially suppressed by sampling \cite{carraro_sampling_2022,carraro_debiased_2020} or by using popularity-stratified recall \cite{steck_item_2011}, which gives a higher reward for recommending less popular items. The unbiased evaluation method for simulating bandit algorithms introduced in \cite{unbiased_offline_evaluation} was experimentally verified in \cite{reseach_papers_recommendation_evaluation} and found to work only for Top-1 recommendations. 

Regarding time-aware validation processes, \cite{zhao_categorical-attributes-based_2018} proposed leave-last-one-out-cross-validation (LLOOCV), where the validation set only considers the latest iteration of each user.  Although there is a clear past-future distinction for each user, in LLOOCV there is no such distinction between users. \cite{jeunen_fair_2018}  offered a more comprehensive solution called $k$-fold Sliding Window Evaluation (SW-EVAL), which analyzes the recommendation algorithm based only on interactions that occurred after the conclusion of the training data.

Unfortunately, despite the significant progress made by the RS community to debiasing offline evaluation, the techniques often do not adequately reflect the nature of \emph{online} recommender systems. An optimal evaluation technique should review the entire system and consider the platform's high-level objectives, such as the number of clicks on items, the number of items purchased, the number of adverts viewed, the customer lifetime value, etc. \cite{rehorek_manipulating_2019}. Such metrics are unsuitable for the offline setting because a RS's environment is domain-dependent, dynamic, and constantly changing. Furthermore, the results are frequently unreproducible because they are based on the monitoring of implicit interactions of a real user or asking them to use pop-up surveys for explicit ratings of the recommended items \cite{canamares_offline_2020}.

This paper fulfills the research objectives outlined for the EvalRS 2023 workshop \cite{bianchi2023evalrs} by experimentally exploring the relationship between offline and online metrics. We note that this comparison type is crucial for real-world recommenders' commercial success. In real scenarios, it is possible to compare a large number of models by using offline metrics. However, to avoid compromising the system's accuracy, only a limited number of models can be examined online. Our first step is to investigate which offline evaluation criteria enables to navigate towards architectures and parametrizations with better online performance. More specifically, we are interested in verifying whether a model with the highest Recall@$k$ (when evaluated offline) also produces the highest CTR (when evaluated online). We also verified the impact of including popularity-penalization and time aspect on the CTR-Recall@$k$ relation. Finally, we introduce a new offline evaluation metric more adaptive to live environments. Our metric, so-called $recall@K_{LLOO}^{\beta}$, simultaneously incorporates popularity penalization and time dependency of interactions.

\section{Related work}
Due to the complexity of performing online evaluations, comparative studies between online and offline metrics are scarce. Added to that, existing solutions typically analyze a single dataset. For instance, in \cite{garcin_offline_2014} the authors compare offline and online metrics on the Swiss news website \textit{swissinfo.ch}. They constructed a metric titled Success@$k$ and used it to fit the model. The value of Success@$k$ in the (offline) validation set is then compared to the CTR in the online environment. As a result, they hypothesize that RAs that are dominant based on offline metrics are no longer as effective in the online context since they favor popular items. In contrast, the RA that recommends random items does significantly better in the online evaluation than in the offline evaluation because they encourage content exploration. As a follow-up, in \cite{predicting_online_performance_swissinfo} they proposed a model predicting online performance based on five different offline metrics but did not find a universal formula for predicting online performance. A similar mismatch between offline and online evaluation is shown for individual domains. For example, in \cite{rossetti_contrasting_2016}, the authors used the MovieLens dataset and evaluated the online performance of RA based on feedback from 100 users invited to their experiment. The behavior of Docear users receiving recommendations of research articles was investigated in \cite{beel_comparative_2013, reseach_papers_recommendation_evaluation} with the conclusion that offline evaluations are probably not suitable to evaluate recommender systems in this domain. An experiment involving 4287 users of a travel agency was conducted in \cite{peska_off-line_2020}. However, according to the findings reported in \cite{jeunen_revisiting_2019}, a general cross-domain comparison is missing. A study from Netflix \cite{steck_deep_2021} from 2021 describes the same issue regarding deep-learning models. They identify a "mismatch in offline and online settings" as one of the unresolved practical challenges of current recommendation systems. The authors of the study give a hint that Netflix uses its own offline bias-suppressing metric for more corresponding offline evaluation. However, they failed to describe essential details, such as how to use contextual bandit techniques to remove various biases. The researchers in \cite{reclist} addressed the challenge of offline evaluation in industry recommender systems by investigating behavioral principles and developing RecList, an open-source tool that employs NLP techniques to assess the effectiveness of a RA in common recommendation scenarios. 
All of the previously referenced research compare offline to online metrics without taking popularity bias and time-dependence into account. Differently, we investigate how a correction for popularity bias~\cite{steck_item_2011} and a validation set consisting purely of interactions that happened after the end of the training data~\cite{jeunen_fair_2018} affect the optimization of online metrics. We also analyzed multiple datasets.

\subsection{Contributions}

The main contributions of this paper are listed as follows:
\begin{itemize}
    \item We introduced a new offline evaluation criterion that included both (1) popularity penalization and (2) time aspect of interactions ($recall@K_{LLOO}^{\beta}$, see Section~\ref{sec:offline-metrics}).
    \item We conducted a large-scale experiment on real-world datasets that shows our criterion has an advantage in predicting online performance over the popular criteria like recall@N used in the RecSys community.
\end{itemize}

\section{Method}
In our experiment, we are examining whether (1) incorporating a time dimension into an offline evaluation approach, and (2) reducing popularity bias by assigning less weight to errors of frequently interacted items, reduces the disparity between offline and online metrics.  In theory, one expects that offline metrics (when employed to cross-validate models) result in improved online performance in live environments.

\noindent \textbf{Basic notation:} Denote the set of interactions between users and items by $F = \{f_1, \dots, f_p\}$.
A single interaction is denoted as $f_j \in (U \times I \times \mathbb{Z}_t) \text{ for } \forall j \in \{1, \dots, p\}$,
 where $U = \{u_1, \dots, u_m\}$ is a set of all RS users, $I = \{i_1, \dots, i_n\}$ is a set of items which can be recommended, $\mathbb{Z}_t$ is the set of integers expressing the timestamp when the interaction was performed and $p$ is a total number of interactions. Then we represent the set of interactions between the item $i$ and the user $u$ as 
\begin{equation*}\label{eq:one-interaction-tuple}
    F_{u,i} = \{(u_j, i_j, t_j) \mid (u_j, i_j, t_j) \in F:  u_j = u \land i_j = i\}.
\end{equation*}

Finally, define the set of relevant items for the user $u$ as $N_u \subset I$. We assume that items that are relevant for a given user can be extracted from explicit and/or implicit feedback. Thus, we considered the item $i$ relevant to the user $u$ if $u$ implicitly (e.g., view, click) interacted with $i$.

\subsection{Offline metrics}\label{sec:offline-metrics}

We will first present offline metrics, which will then be compared to online measurements. Regarding the cross-validating split procedure, we are considering LOOCV and LLOOCV. The most common way to compute recall measured using LOOCV is expressed by:

\begin{equation}\label{eq:leave-one-out-recall}
    recall@K_{LOO} = \frac{
        \sum\limits_{u \in U}
        \sum\limits_{i \in N_u}
        \mid \{i\} \cap Top_K(N_u \setminus \{i\})\mid
    }{
        \sum\limits_{u \in U} \mid N_u \mid
    },
\end{equation}

\noindent where $Top_K(M)$ is a recommendation algorithm that selects $K$ items based on items in the set $M \subset I$ recommendable items. Conversely, recall measured using LLOOCV can be computed by

\begin{equation}\label{eq:leave-last-one-out-recall}
    recall@K_{LLOO} = \frac{
        \sum\limits_{u \in U}
        \sum\limits_{(i_1, t_1) \in F_{u}}
        \mid \{i_1\} \cap Top_K(Q_{t_1}) \mid
    }{
        \sum\limits_{u \in U} \mid N_u \mid
    },
\end{equation}
    with 
\begin{equation*}
    Q_{t_1} = \{i_2 \mid (i_2, t_2) \in F_{u}: t_2 < t_1 \} 
\end{equation*}

\noindent where $F_{u}$ is a list of interactions of user $u$ defined as $F_{u} = \{(i_j, t_j) \mid (u_j, i_j, t_j) \in F:  u_j = u\}$ and the expression $\{i_2 \mid (i_2, t_2) \in F_{u}: t_2 < t_1 \}$ represents the set of items that were interacted by the user $u$ before timestamp~$t_1$. 

Including popularity-penalization from \cite{steck_item_2011} and incorporating it into Eq.~\eqref{eq:leave-one-out-recall}, we get
\small
\begin{equation}\label{eq:ps-loo-recall}
        recall@K_{LOO}^{\beta} = \sum\limits_{u \in U}
        w^{\beta}(u) \frac{
            \sum\limits_{i \in N_u} \mid \{i\} \cap Top_K(N_u \setminus \{i\})\mid \; p(i)^{-\beta}
        }{
            \sum\limits_{i \in N_u} p(i)^{-\beta}
        }
\end{equation}
\normalsize 
and Eq.~\eqref{eq:leave-last-one-out-recall} turns to 
\begin{equation}\label{eq:ps-lloo-recall}
    recall@K_{LLOO}^{\beta} = \sum\limits_{u \in U}
    w^{\beta}(u) \frac{
        \sum\limits_{(i_1, t_1) \in F_{u}} 
        \mid \{i_1\} \cap Top_K(Q_{t_1}) \mid \; p(i)^{-\beta}
    }{
        \sum\limits_{i \in N_u} p(i)^{-\beta}
    }
\end{equation}
where $w^\beta (u) \in [0, 1]$ is a weight of user $u$. The sum of weights for all users must sum up to one, $\sum\limits_{u \in U} w^{\beta} = 1$, with suggested:
\begin{equation*}
    w^{\beta}(u) = \frac{
        1
    } {
        \mid U \mid
    }
    \;
    \frac{
        \sum\limits_{i \in N_u}p(i)^{-\beta}
    } {
        \sum\limits_{v \in U}\sum\limits_{i \in N_v}p(i)^{-\beta}
    }
\end{equation*}
and $p(i) \in [0, 1]$ denotes relative popularity of item $i$. Note that for $\beta=0$, Eq.~\eqref{eq:ps-loo-recall} and~\eqref{eq:leave-one-out-recall} are equivalent, and Eq.~\eqref{eq:ps-lloo-recall} with~\eqref{eq:leave-last-one-out-recall} as well.

\noindent \textbf{Remark:} In \cite{steck_item_2011}, the authors explain the effect of the hyperparameter $\beta$ for offline evaluation but no longer show whether penalizing popularity makes recall a more appropriate metric for the online environment. \cite{jeunen_fair_2018} proposed that reducing bias by penalization of popularity $\beta$ can lead to a better offline evaluation, but with no experimental verification. 

\subsection{Online evaluation}
Another approach to evaluating a RA is based on feedback from users interacting with a live recommender system. More specifically, we work on a scenario where items are recommended to a user, and the RS collects the user's reactions to the recommended items.  An example of a reaction is when the user clicks on the recommended item. 

The most widely used online metric (given its universality) is the click-through rate (CTR). CTR can be seen as the ratio between the accepted recommendations and all offered recommendations \cite{beel_comparative_2013}. We consider that the user accepted the recommendation if he clicked on at least one of the recommended items. 

\noindent \textbf{Remark:} A recommender system can observe implicit and explicit CTRs. An explicit CTR is calculated if the RS has clear evidence that a user clicked on a particular item as a result of the recommendation. The implicit CTR can be calculated based on interactions if the RS does not have explicit knowledge that the item has been clicked through as a result of the recommendation.

Formally, $C: Z_t \times U \times \{0, 1\}^{I}$ defines a set of recommendations where each recommendation is represented by a timestamp $t \in Z_t$, user $u \in U$ and by a set of recommended items $I' \subset I$.
When assuming that the set of interactions $F$ contains only "clicked-type" interactions, then the implicit CTR can be calculated as:
\begin{equation}\label{eq:implicit-ctr}
    iCTR(d) = \frac{
        \sum\limits_{(t, u, I') \in C} sgn(\mid I' \cap F_{u}(t, d) \mid)
    }{
        \mid C \mid
    },
\end{equation}
where $F_{u} (t, d)$ is a set of items interacted by the user $u$ within time $[t, t+d]$ defined as
\begin{equation*}
    F_{u} (t, d) = \{i_j \mid (i_j, t_j) \in F_{u}: (t_j >= t) \land (t + d >= t_j)\},
\end{equation*} 
and $d$ is a parameter determining how long after the recommendation the user has to interact with the recommended item to mark the recommendation as successful.

\subsection{Methodology of the experiment}

The goal of our experiment is to find out how different versions of $recall@K^{\beta}_{LOO}$ and $recall@K^{\beta}_{LLOO}$ as validation metric relates to CTR. 

Due to the limited resources to conduct online experiments, to perform our analysis, we first select a backbone algorithm: the item-$k$NN algorithm \cite{nikolakopoulos_trust_2022} with the similarity between items measured by the cosine similarities of latent space embeddings. The latent space embeddings are created using matrix factorization from implicit feedback \cite{hu_collaborative_2008} with different data preprocessing and hyperparameters (such as latent space size and regularization) to ensure diverse performance according to offline metrics. 
Second, we train our model and measure its performance using different versions of recall (i.e., $recall@K^{\beta}_{LOO}$ and $recall@K^{\beta}_{LLOO}$  with $\beta$ as a hyperparameter).
Third, our model is deployed to an RS with live interactions, and thus the CTR can be measured. Note that the RS is constantly receiving new interactions from users. Therefore, we re-trained the model to include them periodically.

Once the individual steps of the experiment are fulfilled and implemented, the experiment is performed on several datasets. The results of the experiment are measured CTRs along with the number of users that interacted with each model. The RS splits users for each model equally. Once the user was assigned to the model during the first recommendation, the same model generated any further recommendations.

Subsequently, the vector of CTRs can be taken for the dataset $A$ and its $L=5$ deployed models $S^A = (S_1^A, S_2^A, \dots, S_5^A)$. Similarly, one $E^A = (E_1^A, E_2^A, \dots, E_5^A)$ vector is measured for each combination of hyperparameters $V\!AL$, $\beta, k$. The entire list of vectors of recalls is measured according to their hyperparameters $EL^A = \{ E^A_{V\!AL_1, \beta_1, k_1}, E^A_{V\!AL_2, \beta_2, k_2}, \dots, E^A_{V\!AL_V, \beta_V, k_V} \}$, where $V$ is the total number of combinations of hyperparameters and $V\!AL_j$, $\beta_j, k_j$ are the individual hyperparameter values.

Inspired by practical application, we are interested in whether the best model according to offline metrics for a given dataset is also the best according to online metrics. In other words, we find out what the chances are that if we choose the best model according to recall, it will be the best model according to CTR. This can be measured using Recall@$1$. To distinguish between $recall@K^{\beta}_{LLOO}$ measuring the quality of item recommendations and Recall@$1$ measuring whether the best model according to offline metrics is also the best model in online evaluation, we will explicitly denote the latter as \textbf{Model Selection Recall (MSR)}. We define \textbf{MSR} as the ratio of how many times the best model according to offline metrics has been selected by offline metrics, namely recalls with different hyperparameters ($\beta$, $V\!AL$).

\section{Experiments}

In this section, we will describe the used datasets and present the results of our large-scale experiment.

\subsection{Datasets and collected data}
Commonly stable research datasets (such as \textit{MovieLens} or \textit{Last.fm}) cannot be used in our experiment since live users are required for online evaluation. Because of that, our work was performed by using real datasets with live customers. For each model and each dataset, iCTR with a parameter $d=10$ (minutes) was measured over 18 days. The number of users and the number of their recommendations participating in iCTR measurement depend on the data set and other external circumstances and are shown in Table~\ref{tab:datasets-description}. 
\begin{table*} 
    \caption{Description of the total number of items and users in the datasets along with the average number of users and interactions per model used to measure iCTR}
    \label{tab:datasets-description}
    \resizebox{\textwidth}{!}{
        \begin{tabular}{l|l|r|r|r|r|r}
             Dataset & Description & Interactions & Users & Items & Users per model & Interactions per model \\
             \hline
             A & Liquor e-shop & 2.8m & 1.3m & 3k & 7.0k & 22.7k \\
             B & Pet store & 16.8m & 6.9m & 20k & 68.8k & 146.3k \\
             C & Fashion e-shop & 30.4m & 3.1m & 19.3k & 6.3k & 19.1k  \\
             D & Supplier of African goods & 13.9m & 70k & 1k &  19.4k & 67.0k \\
             E & Videostreaming service & 30.7m & 1.8m & 5.3k &  19.6k & 69.7k \\
         \end{tabular}
     }
\end{table*}

The datasets have been selected to include different domains. In addition to e-commerce services, a video streaming platform is represented. Also, dataset D is very different from other e-commerce customers as it is a B2B business with a few products and customers with an unconventional high number of interactions. Another parameter by which the datasets were selected is the number of recommendations per day. The datasets with small traffic could not be selected since the traffic needs to be divided between models and an estimate of CTR needs to be as accurate as possible.

\subsection{Used hyperparameters and resulting MSR
}

\begin{figure}
    \caption{Including a time aspect in the evaluation using the LLOCV approach increases the chance that the selected model will be the best online model (measured by MSR)}
    \label{fig:ctr-recall-comparison}
    \centering
    \includegraphics[width=0.6\linewidth]{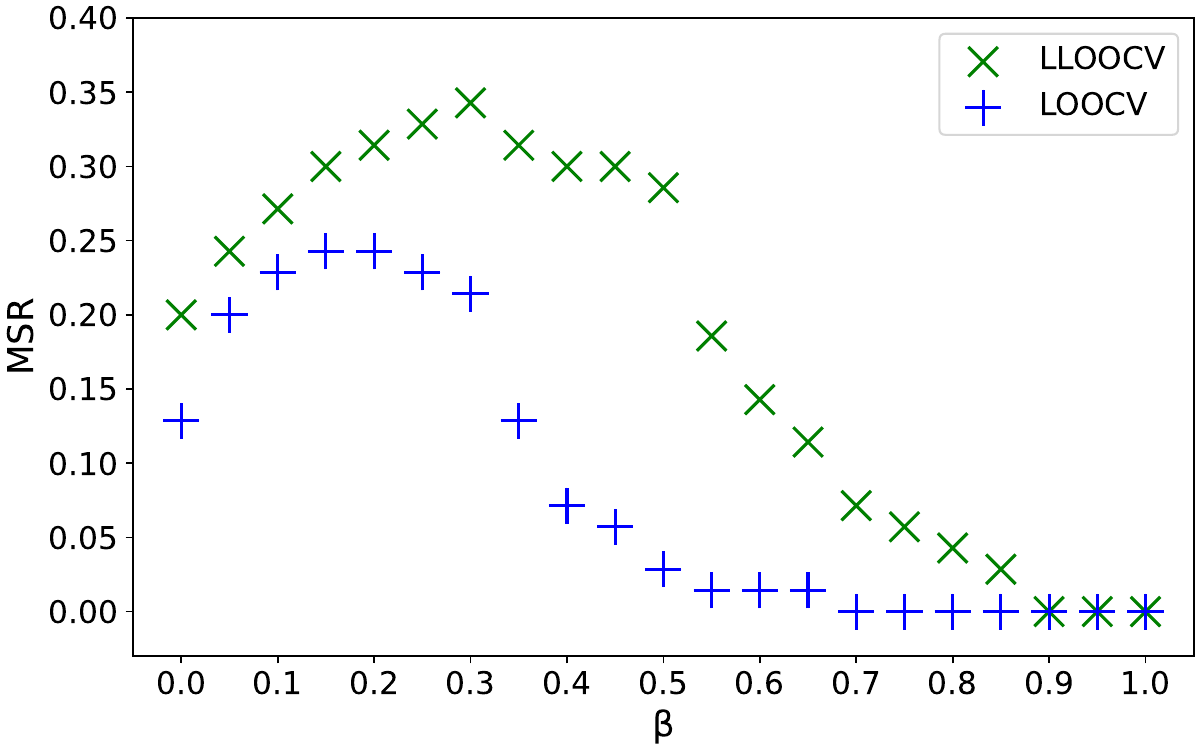}
\end{figure}

The values of hyperparameters for which recall was measured were $k \in \{1,  2,  3,  4,  5,  6,  7,  8,  9, 10, 15, 20, 25, 50\}$ and $\beta \in \{0.05j \mid j \in \{0, 1, \dots, 20\}\}$.
The $V\!AL$ hyperparameter specifies whether is used Equation~\eqref{eq:ps-loo-recall}~or~\eqref{eq:ps-lloo-recall} for the cross-validation method. Figure~\ref{fig:ctr-recall-comparison} shows the relation between iCTR and recall for selected $\beta$s and both values of $V\!AL$ hyperparameter. It can be seen that the inclusion of the time aspect using the LLOCV technique led to an increase in MSR, which means the best online model was more often selected using the offline metric. The same effect can be seen when using the penalization of popularity by the beta hyperparameter. However, it is not true that the more popularity is penalized, the better the offline metric is. 

The best metric found with LLOOCV and $\beta=0.30$ has MSR$=34.29\%$ and therefore, it is better in choosing the best online model than commonly used LOOCV technique with $\beta=0$ (i.e., without popularity penalization) with MSR$=12.86\%$.

\section{Conclusion}

We researched various solutions to better address the known issues of offline metrics with the aim of reducing the gap between offline and online metrics. Then we performed a large-scale experiment that examined the effect of penalizing popularity and including the time aspect of recommendations on the relation between recall and CTR. The experiment was conducted on five commercial datasets covering multiple domains (such as e-commerce, video streaming). It was shown that, in general, maximizing recall does not always lead to maximizing CTR. Popularity-stratified recall proved to be a good approach for reducing the MNAR problem leading to better model selection for an online environment but measured using offline metrics. Offline evaluation with an emphasis on the time aspect of recommendations using leave-last-one-out cross-validation also showed that it can lead to an improvement in MSR. Using the proposed method, we were able to improve the selection of the best online model from the original MSR$ = 12.86\%$ to MSR$ = 34.29\%$.

\subsection{Future work}
The relation between recall and CTR was investigated on five datasets and each contained five different models. To measure the relation more accurately, it would be advisable to have more models, but this greatly prolongs the experiment (since there is a limit traffic of new users) and increases the already exhaustive engineering effort. Similarly, it would be interesting to include datasets from other domains, such as news or bazaars, as these are very specific to the changing popularity of the item. 
As a final improvement, the SW-EVAL method could be used for stricter adherence to the time aspect. Another extension advisable for the future is to examine how the best possible offline metric relates to dataset attributes such as number of users, number of items, sparsity of the interaction matrix, and domain of the dataset. This requires a large meta-study based on experiments with dozens of real datasets for a long period of time.

\begin{acknowledgments}
  This work was supported by the Grant Agency of the Czech Technical University in Prague (SGS23/210/OHK3/3T/18). The experiment was carried out in cooperation with Recombee.
\end{acknowledgments}

\bibliography{manuscript}

\begin{thebibliography}{34}
\expandafter\ifx\csname natexlab\endcsname\relax\def\natexlab#1{#1}\fi
\providecommand{\url}[1]{\texttt{#1}}
\providecommand{\href}[2]{#2}
\providecommand{\path}[1]{#1}
\providecommand{\DOIprefix}{doi:}
\providecommand{\ArXivprefix}{arXiv:}
\providecommand{\URLprefix}{URL: }
\providecommand{\Pubmedprefix}{pmid:}
\providecommand{\doi}[1]{\href{http://dx.doi.org/#1}{\path{#1}}}
\providecommand{\Pubmed}[1]{\href{pmid:#1}{\path{#1}}}
\providecommand{\bibinfo}[2]{#2}
\ifx\xfnm\relax \def\xfnm[#1]{\unskip,\space#1}\fi
\bibitem[{Ledent et~al.(2021)Ledent, Alves, and Kloft}]{ledent_orthogonal_2021}
\bibinfo{author}{A.~Ledent}, \bibinfo{author}{R.~Alves},
  \bibinfo{author}{M.~Kloft},
\newblock \bibinfo{title}{Orthogonal {Inductive} {Matrix} {Completion}},
\newblock \bibinfo{journal}{IEEE Transactions on Neural Networks and Learning
  Systems}  (\bibinfo{year}{2021}) \bibinfo{pages}{1--12}.
  \DOIprefix\doi{10.1109/TNNLS.2021.3106155}, \bibinfo{note}{arXiv:2004.01653
  [cs, stat]}.
\bibitem[{Silveira et~al.(2019)Silveira, Zhang, Lin, Liu, and
  Ma}]{silveira_how_2019}
\bibinfo{author}{T.~Silveira}, \bibinfo{author}{M.~Zhang},
  \bibinfo{author}{X.~Lin}, \bibinfo{author}{Y.~Liu}, \bibinfo{author}{S.~Ma},
\newblock \bibinfo{title}{How good your recommender system is? {A} survey on
  evaluations in recommendation},
\newblock \bibinfo{journal}{International Journal of Machine Learning and
  Cybernetics} \bibinfo{volume}{10} (\bibinfo{year}{2019})
  \bibinfo{pages}{813--831}. \DOIprefix\doi{10.1007/s13042-017-0762-9}.
\bibitem[{Ni et~al.(2019)Ni, Li, and McAuley}]{ni_justifying_2019}
\bibinfo{author}{J.~Ni}, \bibinfo{author}{J.~Li}, \bibinfo{author}{J.~McAuley},
\newblock \bibinfo{title}{Justifying {Recommendations} using
  {Distantly}-{Labeled} {Reviews} and {Fine}-{Grained} {Aspects}},
\newblock in: \bibinfo{booktitle}{Proceedings of the 2019 {Conference} on
  {Empirical} {Methods} in {Natural} {Language} {Processing} and the 9th
  {International} {Joint} {Conference} on {Natural} {Language} {Processing}
  ({EMNLP}-{IJCNLP})}, \bibinfo{publisher}{Association for Computational
  Linguistics}, \bibinfo{address}{Hong Kong, China}, \bibinfo{year}{2019}, pp.
  \bibinfo{pages}{188--197}. \URLprefix
  \url{https://www.aclweb.org/anthology/D19-1018}.
  \DOIprefix\doi{10.18653/v1/D19-1018}.
\bibitem[{Wannigamage et~al.(2020)Wannigamage, Barlow, Lakshika, and
  Kasmarik}]{wannigamage_steam_2020}
\bibinfo{author}{D.~Wannigamage}, \bibinfo{author}{M.~Barlow},
  \bibinfo{author}{E.~Lakshika}, \bibinfo{author}{K.~Kasmarik},
\newblock \bibinfo{title}{Steam {Games} {Dataset} : {Player} count history,
  {Price} history and data about games},
\newblock volume~\bibinfo{volume}{1}, \bibinfo{year}{2020}. \URLprefix
  \url{https://data.mendeley.com/datasets/ycy3sy3vj2/1}.
  \DOIprefix\doi{10.17632/ycy3sy3vj2.1}, \bibinfo{note}{publisher: Mendeley
  Data}.
\bibitem[{Gunawardana et~al.(2022)Gunawardana, Shani, and
  Yogev}]{handbook-evaluating-recommender-systems}
\bibinfo{author}{A.~Gunawardana}, \bibinfo{author}{G.~Shani},
  \bibinfo{author}{S.~Yogev}, \bibinfo{title}{Evaluating Recommender Systems},
  \bibinfo{publisher}{Springer US}, \bibinfo{address}{New York, NY},
  \bibinfo{year}{2022}, pp. \bibinfo{pages}{547--601}. \URLprefix
  \url{https://doi.org/10.1007/978-1-0716-2197-4_15}.
  \DOIprefix\doi{10.1007/978-1-0716-2197-4_15}.
\bibitem[{Wang et~al.(2019)Wang, Zhang, Zhang, Leskovec, Zhao, Li, and
  Wang}]{wang_knowledge-aware_2019}
\bibinfo{author}{H.~Wang}, \bibinfo{author}{F.~Zhang},
  \bibinfo{author}{M.~Zhang}, \bibinfo{author}{J.~Leskovec},
  \bibinfo{author}{M.~Zhao}, \bibinfo{author}{W.~Li},
  \bibinfo{author}{Z.~Wang},
\newblock \bibinfo{title}{Knowledge-aware {Graph} {Neural} {Networks} with
  {Label} {Smoothness} {Regularization} for {Recommender} {Systems}},
\newblock in: \bibinfo{booktitle}{Proceedings of the 25th {ACM} {SIGKDD}
  {International} {Conference} on {Knowledge} {Discovery} \& {Data} {Mining}},
  {KDD} '19, \bibinfo{publisher}{Association for Computing Machinery},
  \bibinfo{address}{New York, NY, USA}, \bibinfo{year}{2019}, pp.
  \bibinfo{pages}{968--977}. \DOIprefix\doi{10.1145/3292500.3330836}.
\bibitem[{Sachdeva et~al.(2018)Sachdeva, Manco, Ritacco, and
  Pudi}]{sachdeva_sequential_2018}
\bibinfo{author}{N.~Sachdeva}, \bibinfo{author}{G.~Manco},
  \bibinfo{author}{E.~Ritacco}, \bibinfo{author}{V.~Pudi},
  \bibinfo{title}{Sequential {Variational} {Autoencoders} for {Collaborative}
  {Filtering}}, \bibinfo{type}{Technical Report}
  \bibinfo{number}{arXiv:1811.09975}, arXiv, \bibinfo{year}{2018}.
  \bibinfo{note}{ArXiv:1811.09975 [cs, stat] type: article}.
\bibitem[{Steck(2019)}]{steck_embarrassingly_2019}
\bibinfo{author}{H.~Steck},
\newblock \bibinfo{title}{Embarrassingly {Shallow} {Autoencoders} for {Sparse}
  {Data}},
\newblock in: \bibinfo{booktitle}{The {World} {Wide} {Web} {Conference}}, {WWW}
  '19, \bibinfo{publisher}{Association for Computing Machinery},
  \bibinfo{address}{New York, NY, USA}, \bibinfo{year}{2019}, pp.
  \bibinfo{pages}{3251--3257}. \DOIprefix\doi{10.1145/3308558.3313710}.
\bibitem[{Yang et~al.(2017)Yang, Bai, Zhang, Yuan, and
  Han}]{yang_bridging_2017}
\bibinfo{author}{C.~Yang}, \bibinfo{author}{L.~Bai},
  \bibinfo{author}{C.~Zhang}, \bibinfo{author}{Q.~Yuan},
  \bibinfo{author}{J.~Han},
\newblock \bibinfo{title}{Bridging {Collaborative} {Filtering} and
  {Semi}-{Supervised} {Learning}: {A} {Neural} {Approach} for {POI}
  {Recommendation}},
\newblock in: \bibinfo{booktitle}{Proceedings of the 23rd {ACM} {SIGKDD}
  {International} {Conference} on {Knowledge} {Discovery} and {Data} {Mining}},
  {KDD} '17, \bibinfo{publisher}{Association for Computing Machinery},
  \bibinfo{address}{New York, NY, USA}, \bibinfo{year}{2017}, pp.
  \bibinfo{pages}{1245--1254}. \URLprefix
  \url{https://doi.org/10.1145/3097983.3098094}.
  \DOIprefix\doi{10.1145/3097983.3098094}.
\bibitem[{Tay et~al.(2018)Tay, Anh~Tuan, and Hui}]{tay_latent_2018}
\bibinfo{author}{Y.~Tay}, \bibinfo{author}{L.~Anh~Tuan}, \bibinfo{author}{S.~C.
  Hui},
\newblock \bibinfo{title}{Latent {Relational} {Metric} {Learning} via
  {Memory}-based {Attention} for {Collaborative} {Ranking}},
\newblock in: \bibinfo{booktitle}{Proceedings of the 2018 {World} {Wide} {Web}
  {Conference}}, {WWW} '18, \bibinfo{publisher}{International World Wide Web
  Conferences Steering Committee}, \bibinfo{address}{Republic and Canton of
  Geneva, CHE}, \bibinfo{year}{2018}, pp. \bibinfo{pages}{729--739}.
  \DOIprefix\doi{10.1145/3178876.3186154}.
\bibitem[{Marlin et~al.(2007)Marlin, Zemel, Roweis, and
  Slaney}]{marlin_collaborative_2007}
\bibinfo{author}{B.~M. Marlin}, \bibinfo{author}{R.~S. Zemel},
  \bibinfo{author}{S.~Roweis}, \bibinfo{author}{M.~Slaney},
\newblock \bibinfo{title}{Collaborative filtering and the missing at random
  assumption},
\newblock in: \bibinfo{booktitle}{Proceedings of the {Twenty}-{Third}
  {Conference} on {Uncertainty} in {Artificial} {Intelligence}}, {UAI}'07,
  \bibinfo{publisher}{AUAI Press}, \bibinfo{address}{Arlington, Virginia, USA},
  \bibinfo{year}{2007}, pp. \bibinfo{pages}{267--275}.
\bibitem[{Marlin and Zemel(2009)}]{marlin_collaborative_2009}
\bibinfo{author}{B.~M. Marlin}, \bibinfo{author}{R.~S. Zemel},
\newblock \bibinfo{title}{Collaborative prediction and ranking with non-random
  missing data},
\newblock in: \bibinfo{booktitle}{Proceedings of the third {ACM} conference on
  {Recommender} systems}, {RecSys} '09, \bibinfo{publisher}{Association for
  Computing Machinery}, \bibinfo{address}{New York, NY, USA},
  \bibinfo{year}{2009}, pp. \bibinfo{pages}{5--12}.
  \DOIprefix\doi{10.1145/1639714.1639717}.
\bibitem[{Steck(2011)}]{steck_item_2011}
\bibinfo{author}{H.~Steck},
\newblock \bibinfo{title}{Item popularity and recommendation accuracy},
\newblock in: \bibinfo{booktitle}{Proceedings of the fifth {ACM} conference on
  {Recommender} systems}, {RecSys} '11, \bibinfo{publisher}{Association for
  Computing Machinery}, \bibinfo{address}{New York, NY, USA},
  \bibinfo{year}{2011}, pp. \bibinfo{pages}{125--132}.
  \DOIprefix\doi{10.1145/2043932.2043957}.
\bibitem[{Li and She(2017)}]{li_collaborative_2017}
\bibinfo{author}{X.~Li}, \bibinfo{author}{J.~She},
\newblock \bibinfo{title}{Collaborative {Variational} {Autoencoder} for
  {Recommender} {Systems}},
\newblock in: \bibinfo{booktitle}{Proceedings of the 23rd {ACM} {SIGKDD}
  {International} {Conference} on {Knowledge} {Discovery} and {Data} {Mining}},
  \bibinfo{publisher}{ACM}, \bibinfo{address}{Halifax NS Canada},
  \bibinfo{year}{2017}, pp. \bibinfo{pages}{305--314}.
  \DOIprefix\doi{10.1145/3097983.3098077}.
\bibitem[{Vinagre et~al.(2014)Vinagre, Jorge, and
  Gama}]{vinagre_evaluation_2014}
\bibinfo{author}{J.~Vinagre}, \bibinfo{author}{A.~M. Jorge},
  \bibinfo{author}{J.~Gama},
\newblock \bibinfo{title}{Evaluation of recommender systems in streaming
  environments},
\newblock \bibinfo{address}{Silicon Valley, United States},
  \bibinfo{year}{2014}. \DOIprefix\doi{10.13140/2.1.4381.5367},
  \bibinfo{note}{arXiv:1504.08175 [cs]}.
\bibitem[{Carraro and Bridge(2022)}]{carraro_sampling_2022}
\bibinfo{author}{D.~Carraro}, \bibinfo{author}{D.~Bridge},
\newblock \bibinfo{title}{A sampling approach to {Debiasing} the offline
  evaluation of recommender systems},
\newblock \bibinfo{journal}{Journal of Intelligent Information Systems}
  \bibinfo{volume}{58} (\bibinfo{year}{2022}) \bibinfo{pages}{311--336}.
  \URLprefix \url{https://link.springer.com/10.1007/s10844-021-00651-y}.
  \DOIprefix\doi{10.1007/s10844-021-00651-y}.
\bibitem[{Carraro and Bridge(2020)}]{carraro_debiased_2020}
\bibinfo{author}{D.~Carraro}, \bibinfo{author}{D.~Bridge},
\newblock \bibinfo{title}{Debiased offline evaluation of recommender systems: a
  weighted-sampling approach},
\newblock in: \bibinfo{booktitle}{Proceedings of the 35th {Annual} {ACM}
  {Symposium} on {Applied} {Computing}}, {SAC} '20,
  \bibinfo{publisher}{Association for Computing Machinery},
  \bibinfo{address}{New York, NY, USA}, \bibinfo{year}{2020}, pp.
  \bibinfo{pages}{1435--1442}. \URLprefix
  \url{https://doi.org/10.1145/3341105.3375759}.
  \DOIprefix\doi{10.1145/3341105.3375759}.
\bibitem[{Li et~al.(2011)Li, Chu, Langford, and
  Wang}]{unbiased_offline_evaluation}
\bibinfo{author}{L.~Li}, \bibinfo{author}{W.~Chu},
  \bibinfo{author}{J.~Langford}, \bibinfo{author}{X.~Wang},
\newblock \bibinfo{title}{Unbiased offline evaluation of
  contextual-bandit-based news article recommendation algorithms},
\newblock in: \bibinfo{booktitle}{Proceedings of the Fourth ACM International
  Conference on Web Search and Data Mining}, WSDM '11,
  \bibinfo{publisher}{Association for Computing Machinery},
  \bibinfo{address}{New York, NY, USA}, \bibinfo{year}{2011}, p.
  \bibinfo{pages}{297–306}. \URLprefix
  \url{https://doi.org/10.1145/1935826.1935878}.
  \DOIprefix\doi{10.1145/1935826.1935878}.
\bibitem[{Beel and Langer(2015)}]{reseach_papers_recommendation_evaluation}
\bibinfo{author}{J.~Beel}, \bibinfo{author}{S.~Langer},
\newblock \bibinfo{title}{A comparison of offline evaluations, online
  evaluations, and user studies in the context of research-paper recommender
  systems},
\newblock in: \bibinfo{editor}{S.~Kapidakis}, \bibinfo{editor}{C.~Mazurek},
  \bibinfo{editor}{M.~Werla} (Eds.), \bibinfo{booktitle}{Research and Advanced
  Technology for Digital Libraries}, \bibinfo{publisher}{Springer International
  Publishing}, \bibinfo{address}{Cham}, \bibinfo{year}{2015}, pp.
  \bibinfo{pages}{153--168}.
\bibitem[{Zhao et~al.(2018)Zhao, Chen, Chen, Jain, Beutel, Belletti, and
  Chi}]{zhao_categorical-attributes-based_2018}
\bibinfo{author}{Q.~Zhao}, \bibinfo{author}{J.~Chen},
  \bibinfo{author}{M.~Chen}, \bibinfo{author}{S.~Jain},
  \bibinfo{author}{A.~Beutel}, \bibinfo{author}{F.~Belletti},
  \bibinfo{author}{E.~H. Chi},
\newblock \bibinfo{title}{Categorical-attributes-based item classification for
  recommender systems},
\newblock in: \bibinfo{booktitle}{Proceedings of the 12th {ACM} {Conference} on
  {Recommender} {Systems}}, {RecSys} '18, \bibinfo{publisher}{Association for
  Computing Machinery}, \bibinfo{address}{New York, NY, USA},
  \bibinfo{year}{2018}, pp. \bibinfo{pages}{320--328}.
  \DOIprefix\doi{10.1145/3240323.3240367}.
\bibitem[{Jeunen et~al.(2018)Jeunen, Verstrepen, and
  Goethals}]{jeunen_fair_2018}
\bibinfo{author}{O.~Jeunen}, \bibinfo{author}{K.~Verstrepen},
  \bibinfo{author}{B.~Goethals},
\newblock \bibinfo{title}{Fair {Offline} {Evaluation} {Methodologies} for
  {Implicit}-{Feedback} {Recommender} {Systems} with {MNAR} {Data}},
\newblock \bibinfo{address}{Vancouver, Canada}, \bibinfo{year}{2018},
  p.~\bibinfo{pages}{8}. \URLprefix
  \url{http://adrem.uantwerpen.be/bibrem/pubs/OfflineEvalJeunen2018.pdf}.
\bibitem[{Řehořek(2019)}]{rehorek_manipulating_2019}
\bibinfo{author}{T.~Řehořek}, \bibinfo{title}{Manipulating the {Capacity} of
  {Recommendation} {Models} in {Recall}-{Coverage} {Optimization}}, Ph.D.
  thesis, Czech Technical University, \bibinfo{year}{2019}. \URLprefix
  \url{https://dspace.cvut.cz/handle/10467/81823}, \bibinfo{note}{accepted:
  2019-04-05T11:19:10Z Publisher: České vysoké učení technické v Praze.
  Vypočetní a informační centrum.}
\bibitem[{Cañamares et~al.(2020)Cañamares, Castells, and
  Moffat}]{canamares_offline_2020}
\bibinfo{author}{R.~Cañamares}, \bibinfo{author}{P.~Castells},
  \bibinfo{author}{A.~Moffat},
\newblock \bibinfo{title}{Offline evaluation options for recommender systems},
\newblock \bibinfo{journal}{Information Retrieval Journal} \bibinfo{volume}{23}
  (\bibinfo{year}{2020}) \bibinfo{pages}{387--410}.
  \DOIprefix\doi{10.1007/s10791-020-09371-3}.
\bibitem[{Bianchi et~al.(2023)Bianchi, Chia, Greco, Pomo, Moreira, Eynard,
  Husain, and Tagliabue}]{bianchi2023evalrs}
\bibinfo{author}{F.~Bianchi}, \bibinfo{author}{P.~J. Chia},
  \bibinfo{author}{C.~Greco}, \bibinfo{author}{C.~Pomo},
  \bibinfo{author}{G.~Moreira}, \bibinfo{author}{D.~Eynard},
  \bibinfo{author}{F.~Husain}, \bibinfo{author}{J.~Tagliabue},
  \bibinfo{title}{Evalrs 2023. well-rounded recommender systems for real-world
  deployments}, \bibinfo{year}{2023}.
  \href{http://arxiv.org/abs/2304.07145}{{\tt arXiv:2304.07145}}.
\bibitem[{Garcin et~al.(2014)Garcin, Faltings, Donatsch, Alazzawi, Bruttin, and
  Huber}]{garcin_offline_2014}
\bibinfo{author}{F.~Garcin}, \bibinfo{author}{B.~Faltings},
  \bibinfo{author}{O.~Donatsch}, \bibinfo{author}{A.~Alazzawi},
  \bibinfo{author}{C.~Bruttin}, \bibinfo{author}{A.~Huber},
\newblock \bibinfo{title}{Offline and online evaluation of news recommender
  systems at swissinfo.ch},
\newblock in: \bibinfo{booktitle}{Proceedings of the 8th {ACM} {Conference} on
  {Recommender} systems - {RecSys} '14}, \bibinfo{publisher}{ACM Press},
  \bibinfo{address}{Foster City, Silicon Valley, California, USA},
  \bibinfo{year}{2014}, pp. \bibinfo{pages}{169--176}.
  \DOIprefix\doi{10.1145/2645710.2645745}.
\bibitem[{Maksai et~al.(2015)Maksai, Garcin, and
  Faltings}]{predicting_online_performance_swissinfo}
\bibinfo{author}{A.~Maksai}, \bibinfo{author}{F.~Garcin},
  \bibinfo{author}{B.~Faltings},
\newblock \bibinfo{title}{Predicting online performance of news recommender
  systems through richer evaluation metrics},
\newblock in: \bibinfo{booktitle}{Proceedings of the 9th ACM Conference on
  Recommender Systems}, RecSys '15, \bibinfo{publisher}{Association for
  Computing Machinery}, \bibinfo{address}{New York, NY, USA},
  \bibinfo{year}{2015}, p. \bibinfo{pages}{179–186}. \URLprefix
  \url{https://doi.org/10.1145/2792838.2800184}.
  \DOIprefix\doi{10.1145/2792838.2800184}.
\bibitem[{Rossetti et~al.(2016)Rossetti, Stella, and
  Zanker}]{rossetti_contrasting_2016}
\bibinfo{author}{M.~Rossetti}, \bibinfo{author}{F.~Stella},
  \bibinfo{author}{M.~Zanker},
\newblock \bibinfo{title}{Contrasting {Offline} and {Online} {Results} when
  {Evaluating} {Recommendation} {Algorithms}},
\newblock in: \bibinfo{booktitle}{Proceedings of the 10th {ACM} {Conference} on
  {Recommender} {Systems}}, {RecSys} '16, \bibinfo{publisher}{Association for
  Computing Machinery}, \bibinfo{address}{New York, NY, USA},
  \bibinfo{year}{2016}, pp. \bibinfo{pages}{31--34}. \URLprefix
  \url{https://doi.org/10.1145/2959100.2959176}.
  \DOIprefix\doi{10.1145/2959100.2959176}.
\bibitem[{Beel et~al.(2013)Beel, Genzmehr, Langer, Nürnberger, and
  Gipp}]{beel_comparative_2013}
\bibinfo{author}{J.~Beel}, \bibinfo{author}{M.~Genzmehr},
  \bibinfo{author}{S.~Langer}, \bibinfo{author}{A.~Nürnberger},
  \bibinfo{author}{B.~Gipp},
\newblock \bibinfo{title}{A comparative analysis of offline and online
  evaluations and discussion of research paper recommender system evaluation},
\newblock in: \bibinfo{booktitle}{Proceedings of the {International} {Workshop}
  on {Reproducibility} and {Replication} in {Recommender} {Systems}
  {Evaluation}}, {RepSys} '13, \bibinfo{publisher}{Association for Computing
  Machinery}, \bibinfo{address}{New York, NY, USA}, \bibinfo{year}{2013}, pp.
  \bibinfo{pages}{7--14}. \DOIprefix\doi{10.1145/2532508.2532511}.
\bibitem[{Peska and Vojtas(2020)}]{peska_off-line_2020}
\bibinfo{author}{L.~Peska}, \bibinfo{author}{P.~Vojtas},
\newblock \bibinfo{title}{Off-line vs. {On}-line {Evaluation} of {Recommender}
  {Systems} in {Small} {E}-commerce},
\newblock \bibinfo{journal}{Proceedings of the 31st ACM Conference on Hypertext
  and Social Media}  (\bibinfo{year}{2020}) \bibinfo{pages}{291--300}.
  \DOIprefix\doi{10.1145/3372923.3404781}, \bibinfo{note}{arXiv: 1809.03186}.
\bibitem[{Jeunen(2019)}]{jeunen_revisiting_2019}
\bibinfo{author}{O.~Jeunen},
\newblock \bibinfo{title}{Revisiting offline evaluation for implicit-feedback
  recommender systems},
\newblock in: \bibinfo{booktitle}{Proceedings of the 13th {ACM} {Conference} on
  {Recommender} {Systems}}, {RecSys} '19, \bibinfo{publisher}{Association for
  Computing Machinery}, \bibinfo{address}{New York, NY, USA},
  \bibinfo{year}{2019}, pp. \bibinfo{pages}{596--600}.
  \DOIprefix\doi{10.1145/3298689.3347069}.
\bibitem[{Steck et~al.(2021)Steck, Baltrunas, Elahi, Liang, Raimond, and
  Basilico}]{steck_deep_2021}
\bibinfo{author}{H.~Steck}, \bibinfo{author}{L.~Baltrunas},
  \bibinfo{author}{E.~Elahi}, \bibinfo{author}{D.~Liang},
  \bibinfo{author}{Y.~Raimond}, \bibinfo{author}{J.~Basilico},
\newblock \bibinfo{title}{Deep {Learning} for {Recommender} {Systems}: {A}
  {Netflix} {Case} {Study}},
\newblock \bibinfo{journal}{AI Magazine} \bibinfo{volume}{42}
  (\bibinfo{year}{2021}) \bibinfo{pages}{7--18}. \URLprefix
  \url{https://ojs.aaai.org/index.php/aimagazine/article/view/18140}.
  \DOIprefix\doi{10.1609/aimag.v42i3.18140}, \bibinfo{note}{number: 3}.
\bibitem[{Chia et~al.(2022)Chia, Tagliabue, Bianchi, He, and Ko}]{reclist}
\bibinfo{author}{P.~J. Chia}, \bibinfo{author}{J.~Tagliabue},
  \bibinfo{author}{F.~Bianchi}, \bibinfo{author}{C.~He},
  \bibinfo{author}{B.~Ko},
\newblock \bibinfo{title}{Beyond ndcg: Behavioral testing of recommender
  systems with reclist},
\newblock WWW '22 Companion, \bibinfo{publisher}{Association for Computing
  Machinery}, \bibinfo{address}{New York, NY, USA}, \bibinfo{year}{2022}, p.
  \bibinfo{pages}{99–104}. \URLprefix
  \url{https://doi.org/10.1145/3487553.3524215}.
  \DOIprefix\doi{10.1145/3487553.3524215}.
\bibitem[{Nikolakopoulos et~al.(2022)Nikolakopoulos, Ning, Desrosiers, and
  Karypis}]{nikolakopoulos_trust_2022}
\bibinfo{author}{A.~N. Nikolakopoulos}, \bibinfo{author}{X.~Ning},
  \bibinfo{author}{C.~Desrosiers}, \bibinfo{author}{G.~Karypis},
\newblock \bibinfo{title}{Trust {Your} {Neighbors}: {A} {Comprehensive}
  {Survey} of {Neighborhood}-{Based} {Methods} for {Recommender} {Systems}},
\newblock in: \bibinfo{editor}{F.~Ricci}, \bibinfo{editor}{L.~Rokach},
  \bibinfo{editor}{B.~Shapira} (Eds.), \bibinfo{booktitle}{Recommender
  {Systems} {Handbook}}, \bibinfo{publisher}{Springer US},
  \bibinfo{address}{New York, NY}, \bibinfo{year}{2022}, pp.
  \bibinfo{pages}{39--89}. \URLprefix
  \url{https://doi.org/10.1007/978-1-0716-2197-4_2}.
  \DOIprefix\doi{10.1007/978-1-0716-2197-4_2}.
\bibitem[{Hu et~al.(2008)Hu, Koren, and Volinsky}]{hu_collaborative_2008}
\bibinfo{author}{Y.~Hu}, \bibinfo{author}{Y.~Koren},
  \bibinfo{author}{C.~Volinsky},
\newblock \bibinfo{title}{Collaborative {Filtering} for {Implicit} {Feedback}
  {Datasets}},
\newblock \bibinfo{year}{2008}, pp. \bibinfo{pages}{263--272}.
  \DOIprefix\doi{10.1109/ICDM.2008.22}.

\end{thebibliography}

\end{document}